\documentclass[journal, twoside]{IEEETran}

\usepackage[brazilian]{babel}
\usepackage{fancyhdr} 
\usepackage{lipsum}
\usepackage[utf8]{inputenc}
\usepackage[T1]{fontenc}
\usepackage{graphicx} 
\usepackage{subfig}
\usepackage{nccmath}
\usepackage{color}
\usepackage{cite}

\usepackage[figurename=Fig.]{caption} 
\usepackage[labelsep=period]{caption} 

\ifCLASSINFOpdf
\else
\fi
\begin{document}

\setcounter{page}{1}


\title{Standing Forest Coin (SFC)}

\author{
    Marcelo de A. Borges, 
    Guido L. de S. Filho,
    Cicero Inacio da Silva, 
    Anderson M. P. Barros, \\
    Raul V. B. J. Britto,
    Nivaldo M. de C. Junior and 
    Daniel F. L. de Souza 
    
    \thanks{Marcelo de A. Borges is with GT-V4H at RNP, Rio de Janeiro (e-mail: marceloabreuborges@gmail.com)}
    
    \thanks{Guido L. de S. Filho is with LAVID @ Universidade Federal da Paraiba, João Pessoa, PB, Brazil (e-mail: guido@lavid.ufpb.br)}
    
    \thanks{Cicero Inacio da Silva is with Campus São Paulo at Universidade Federal de São Paulo (Unifesp), São Paulo (e-mail: csilva@weber.ucsd.edu)}
    \thanks{Anderson M. P. Barros is with PPGI at Universidade Federal da Paraiba, João Pessoa, PB, Brazil (e-mail: andersonmarinhobarros@gmail.com)}
    \thanks{Raul V. B. J. Britto is with PPGI at Universidade Federal da Paraiba, João Pessoa, PB, Brazil, e-mail: raulbbritto@gmail.com)}
    \thanks{Nivaldo M. de C. Junior is with Universidade Federal da Paraiba, João Pessoa, PB, Brazil (e-mail: nivaldomcj@gmail.com)}
    \thanks{Daniel F. L. Souza is with Lavid at UFPB, João Pessoa, PB, Brazil (e-mail: daniel@lavid.ufpb.br)}
    \thanks{Trabalho desenvolvido a partir da dissertação de mestrado de Marcelo de Abreu Borges denominada ``A Amazônia e o atual paradoxo brasileiro: contexto histórico e uma proposta alternativa de financiamento para a conservação da floresta. Projeto Floresta 4.0'' \cite{marcelodeabreuborges}, defendida no âmbito do Mestrado Profissional em Análise e Gestão de Políticas Internacionais: Resolução de Conflitos e Cooperação para o Desenvolvimento (MAPI) da PUC Rio sob a orientação do Prof. Dr. Paulo Luiz Moreaux Lavigne Esteves e a partir do projeto Marx Coin, desenvolvido pelos Profs. Brett Stalbaum e Cicero Inacio da Silva na Universidade da Califórnia em 2018}
    \thanks{Manuscript received August 25, 2021; revised August 31, 2021}
}




\markboth{Standing Forest Coin - SFC}{SKM: My SFC article}  
%


\maketitle

\begin{abstract}
This article describes a proposal to create a digital currency that allows the descentralized collection of resources directed to initiatives and activities that aim to protect the Brazilian Amazon ecosystem by using blockchain and digital contracts. In addition to the digital currency, the goal is to design a smart contract based in oracles to ensure credibility and security for investors and donors of financial resources invested in projects within the Standing Forest Coin (SFC - standingforest.org).
\end{abstract}


\IEEEpeerreviewmaketitle

\section{Introdução}

\IEEEPARstart{O}{}surgimento diário de diversas criptomoedas ou criptoativos mostra que a blockchain, ideia iniciada após a crise de 2008 e que visa a eliminação da dependência de instituições centralizadas ou intermediárias, torna-se cada vez mais disseminada e aceita. Cada vez mais o conceito de descentralização e antifragilidade apresentado inicialmente pelo Bitcoin \cite{nakamoto} é amplamente divulgado e proposto, como por exemplo: Ethereum \cite{buterin}, Tether \cite{tether} e outras criptomoedas.

A tecnologia proposta pela blockchain não somente reduz a necessidade de intermediários que garantam a integridade de transações de um sistema proposto, mas também possui mecanismos que dirimem a perda de confiança, credibilidade e possibilidade de fraude nas transações. Além disso, o seu funcionamento ainda reduz custos, tempo e burocracia. Acrescente-se a isso mais transparência, segurança e rastreabilidade.

Com a aplicação de conceitos da blockchain, este artigo sugere a criação de uma criptomoeda e contratos inteligentes que atinja liquidez e repercussão mundial, que pode vir a ter o potencial de gerar novas iniciativas e divulgar toda uma agenda de atividades que ajudem a proteger inicialmente o ecossistema amazônico, mas com abertura para o financiamento de atividades de conservação em outros ecossistemas brasileiros, como a Cerrado e a Caatinga.

De acordo com Lopes \cite{matanativa}, a região amazônica é um dos maiores desafios do mundo na busca pelo desenvolvimento de práticas sustentáveis de produção e consumo. Esse enorme desafio de desenvolvimento social, econômico e ambiental sustentável e a conservação desse bioma, inclui necessariamente a questão da produção sustentável – principalmente a criação de emprego e renda a partir da conservação e/ou exploração regrada dos recursos naturais.

Ainda segundo Lopes \cite{matanativa}, é preciso deixar claro que o grande desafio de mudar o modelo de desenvolvimento regional é político e que a ciência e a tecnologia, por si só, são insuficientes quando sozinhas. No entanto, a ciência e a tecnologia desempenham um papel fundamental no desenvolvimento sustentável da Amazônia, considerando a necessidade de novos conhecimentos para o pleno desenvolvimento das cadeias produtivas, valorizando a biodiversidade e os serviços ambientais dos ecossistemas.

A capacitação tecnológica provou ser uma ferramenta fundamental para manter as economias emergentes de importantes países em desenvolvimento. Nos últimos cinquenta anos, o Brasil tem sido capaz de criar ilhas de excelência em ciência e tecnologia, que são mais semelhantes às dos países desenvolvidos do que as de baixa ou média renda \cite{matanativa}. 

No entanto, as desigualdades regionais históricas, especialmente as da educação, criaram impedimentos limitando drasticamente o uso intensivo de ciência e tecnologia para as economias e o desenvolvimento social das regiões mais pobres e menos favorecidas, incluindo a Amazônia e o Nordeste brasileiro \cite{matanativa}. 

Piccolotto em \cite{piccolotto} defende que a chamada Revolução 4.0, conceito dado pela transformação trazida pelo desenvolvimento de novas tecnologias, pode ser um caminho eficaz para alinhar o desenvolvimento estratégico e sustentável à preservação ambiental. 

Um bom exemplo desse movimento é o conceito de Floresta 4.0 que consiste em uma ideia sobre como novas tecnologias, como a blockchain, estão impactando o setor de maneira direta e adjacente, o que deve ser visto com um olhar positivo sobre todo o sistema: desde as terras usadas para a produção até o uso dos produtos e serviços pelo consumidor final.

Floresta 4.0 significa o uso de novas tecnologias para gestão dos ativos florestais: seu objetivo é trazer mais eficiência para o processo produtivo florestal em todas as suas etapas \cite{jotz}. Isto significa maior nível de automação dos equipamentos, maior geração de telemetrias (dados) para o monitoramento, controle e conectividade para uma tomada de decisão com maior agilidade e assertividade \cite{bforest}.

\section{Trabalhos Relacionados}

Nesta seção serão apresentados trabalhos que tenham temática relacionada com a Standing Forest Coin (SFC). Estes trabalhos, em sua maioria, têm como objetivo empregar o uso de blockchains para gerenciamento de recursos em projetos, seja por rastreamento desses recursos para fiscalização ou apenas para uma administração descentralizada dos pagamentos a esses projetos.

\subsection{Alice}

Alice é um projeto de blockchain filantrópico \cite{alice} baseado na blockchain Ethereum. A intenção dessa rede é que as organizações não governamentais (ONGs) forneçam informações sobre os impactos promovidos pelos seus projetos, assim sendo mais transparentes com seus investidores e doadores.

De acordo com os autores, as ONGs e instituições sem fins lucrativos têm dificuldade de arrecadar recursos financeiros com investidores e doadores para execução de seus projetos, e essa dificuldade se dá pela falta de transparência com a maneira que esses recursos são aplicados em seus projetos. Assim, os interessados em investir perdem a confiança em doar para essas organizações, por não terem uma maneira de mensurar a efetividade da execução desses projetos \cite{alice}.

A proposta da Alice é que as organizações registrem na rede blockchain os projetos sociais com todos os detalhes dos objetivos a serem alcançados pelo projeto, sendo esses objetivos verificados de maneira independente por um terceiro de confiança (oráculo). Dessa maneira, após o aporte inicial para o projeto, os recursos financeiros subsequentes são liberados para a organização apenas se ela puder provar que os objetivos descritos no início do projeto foram alcançados.

Para garantir esses princípios, Alice usa a rede pública Ethereum com contratos inteligentes. Segundo os autores, todos os dados do projeto social, desde os objetivos, recursos operacionais, registros do projeto são adicionados à rede, tornando-os públicos. Dados sensíveis, como informações dos beneficiários, são adicionados em off-chain. Os projetos têm um validador independente (oráculo) que verifica se as organizações alcançaram os objetivos.

\subsection{BndesToken}

O BndesToken é um projeto de blockchain que visa definir um procedimento de rastreamento do uso de recursos públicos em projetos de financiamento do Banco Nacional de Desenvolvimento Econômico e Social do Brasil (BNDES) \cite{wblockchain}. A proposta dos autores é a de trazer transparência para a sociedade sobre os recursos públicos investidos e facilitar o acompanhamento das operações desses projetos.

De acordo com os autores, o emprego do blockchain permite que a sociedade acompanhe de forma transparente e acredite na inviolabilidade dos dados e nos projetos de financiamento que são adicionados ao sistema, sem precisar confiar numa entidade centralizadora. O uso de outras tecnologias como o desenvolvimento de um Sistema de Informação Gerencial (SIG), não garantiria a inviolabilidade em tempo real, além de ser custoso para manter a auditoria e as informações dos projetos \cite{wblockchain}.

Assim, o objetivo do projeto é que cada unidade do BndesToken seja equivalente a um Real (1:1) para facilitar a marcação da moeda nacional brasileira, mantendo-a como um lastro. Os ``tokens'' são emitidos para os projetos financiados no momento da liberação do recurso, podendo ser movimentado pelo mantenedor do projeto, mas devendo ser resgatado posteriormente no BNDES. Por fim, os eventos de transferência de recursos não são automaticamente relacionados com os marcos do projeto de financiamento.

Para a implementação, o BndesToken utiliza a plataforma Ethereum com a concepção de contratos inteligentes. Os contratos desenvolvidos utilizam o padrão ERC-20, escritos na linguagem Solidity. O acompanhamento das etapas de um projeto financiado pelo BndesToken pode ser feito, de acordo com os autores, usando ferramentas próprias para visualização de eventos da blockchain Ethereum, como o EtherScan \cite{choi2018}, mas também foi desenvolvido uma aplicação Web para facilitar o acompanhamento.

\subsection{Aidcoin}

O projeto Aidcoin tem como objetivo definir um ecossistema de serviços de fácil acesso que permita investidores e partes interessadas realizar doações de forma transparente de recursos financeiros para projetos em ascensão, utilizando uma rede blockchain \cite{aidcoin}. Esse projeto foi concebido e é mantido pela Charity Stars, uma empresa de arrecadação de recursos para caridade.

Segundo os autores, os recentes escândalos de corrupção e mau uso de recursos por projetos sem fins lucrativos levaram à diminuição da confiança por parte de investidores e possíveis doadores a projetos de caridade. Dessa forma, os investidores estão exigindo das instituições sem fins lucrativos maior transparência e responsabilidade, e essas organizações estão procurando formas de suprir essa necessidade.

A proposta desse projeto é a de usar contratos inteligentes e criptomoedas para prover essa transparência, pois com o uso de blockchains seria possível obter uma redução de custos operacionais de doação, um aumento da transparência nos processos e uma maior eficiência no gerenciamento dessas doações.

O Aidcoin é baseado na plataforma Ethereum e é mantido por uma criptomoeda própria, o Aidcoin Token. O projeto visou, além de implementar o ecossistema em blockchain, integrar serviços como uma exchange própria para converter outras criptomoedas em Aidcoins e um sistema para gerenciar os projetos e doações de maneira transparente, tanto para a organização quanto para o investidor.

\subsection{Comparativo entre os Trabalhos Relacionados}

Os trabalhos relacionados foram analisados e selecionados pela proximidade de propostas e objetivos com a Standing Forest Coin. Todos os projetos descritos procuram, das mais variadas formas, aumentar a transparência no processo de gerenciamento de recursos de projetos, sejam eles públicos ou privados.

Para proporcionar transparência nas etapas de investimento e administração dos recursos dos projetos, todos os trabalhos utilizam redes blockchain públicas. O uso do Ethereum, de acordo com os autores, propicia a audição das informações dos projetos por qualquer indivíduo, além de permitir a criação de contratos para assegurar quando esses recursos serão liberados para as organizações e projetos.

Como diferença explícita, o Alice e Aidcoin pretendem, por meio de mecanismos como oráculos, relacionar a transferência do recurso pelo marco do projeto, liberando os recursos financeiros assim que é atestado o cumprimento do objetivo do projeto, enquanto o BndesToken não possui tal procedimento.

Todos os trabalhos procuram criar formas de acesso aos dados de maneira intuitiva, com o desenvolvimento de sistemas de internet e painéis próprios para a visualização desses recursos, sem a necessidade de acessar os dados diretamente na rede blockchain.

\section{Definição da Proposta}

Neste trabalho é proposto uma moeda digital que promova inicialmente o financiamento e execução de projetos voltados à proteção do ecossistema amazônico, com mecanismos para assegurar o cumprimento desses projetos.

Para esse trabalho optou-se pelo uso de uma rede blockchain pública ao invés de uma blockchain privada, com o objetivo de tornar todas as transações e informações dos projetos e da moeda transparentes.

Nas blockchains privadas, as transações não são visíveis ao público geral \cite{trailhead}. Quando há a possibilidade de visualização do livro-razão de uma blockchain privada, essa se dá de maneira permissionada, sendo necessário uma autoridade centralizadora para o credenciamento e autorização de usuários \cite{olhardigital}.

Por outro lado, as blockchains públicas permitem que qualquer pessoa transacione nessas redes, e que inclusive visualize as entradas e saídas das movimentações feitas por outros usuários \cite{trailhead}.

Existem diversas redes de blockchain públicas, havendo diferenças sutis entre elas, como por exemplo: o tempo em que uma transação leva para ser adicionada no livro razão, a dificuldade de ``mineração'' de uma página do livro razão e a quantidade e limite de unidades da moeda disponível.

Dentre as blockchains públicas, foi escolhida para este projeto o Ethereum, devido ao tempo de atividade e suas características intrínsecas que permitem a concepção desse projeto. O Ethereum é uma blockchain de código aberto que permite a criação de contratos inteligentes e aplicações descentralizadas, que tem como moeda digital o Ether, que pode ser obtido por meio da mineração ou por transações \cite{buterin}.

Blockchains públicas garantem a visualização de toda e qualquer transação realizada, bastando para isso ter acesso ao livro-razão. Por meio de aplicativos de exploração de blocos, é possível visualizar em tempo real as transações e contratos executados na rede Ethereum. O software mais comum para a exploração de blocos da rede Ethereum é o Etherscan \cite{choi2020}, que funciona direto do navegador de internet. Isso permite que qualquer pessoa consiga visualizar de forma transparente as transações realizadas com a SFC.

Um dos objetivos deste trabalho é permitir que as iniciativas de preservação de terra sejam apoiadas financeiramente de forma ágil, à medida que forem disponibilizados insumos comprovando a execução das etapas desses projetos e consequentemente a continuidade da preservação da floresta. Assim, as frações da moeda serão liberadas para os projetos de conservação de acordo com a conferência das informações fornecidas por meio de um oráculo, aumentando a segurança e a credibilidade nos projetos e iniciativas fomentadas com a criptomoeda.

Os contratos inteligentes, por design, são feitos para serem executados do início ao fim dentro do ambiente da blockchain, não sendo possível acessar diretamente dados externos, sendo necessário para tal a consulta a um oráculo. 

Oráculos, na terminologia blockchain, são serviços que provêm informações do mundo real aos contratos inteligentes, permitindo o uso desses dados para a execução destes contratos, seja tanto para assegurar que um evento ocorreu no mundo real, quanto para validar alguma informação fornecida pelas partes do contrato \cite{bit2me}.

A priori, para a SFC foi definido um oráculo baseado em informações de sistema de geoposicionamento (GPS), imagens aeroespaciais e de satélites. Dadas as coordenadas geográficas fornecidas pelo projeto no momento da sua criação, a liberação das frações da SFC para os projetos dar-se-á pela validação desses dados com o oráculo, para verificar, por exemplo, informações sobre a variação da cobertura vegetal da área do projeto durante um intervalo de tempo a fim de garantir que a área foi de fato protegida pelo projeto.

A SFC tem como característica ser transacionável em mercados financeiros, sendo registrada em corretoras de câmbio internacionais. Assim, o caminho de troca da moeda é claro, sendo possível a troca da mesma por Ether (da blockchain Ethereum), que pode também ser trocada por diversas outras moedas, incluindo por moedas fiduciárias, como o dólar.

Por ser transacionável em mercados financeiros, a proposta da moeda é que o contribuinte da moeda teria um ativo, podendo eventualmente ter um ganho de capital decorrente desse ativo com o decorrer do tempo, tendo também a opção de financiar as atividades de valoração de ecossistemas ameaçados.

\section{Estado Atual de Implementação}

\begin{figure*}[h]
\caption{{} Diagrama de Sequência dos Fluxos Presentes na Implementação da SFC}
\label{fig:SequenceDiagram-Case01}
\begin{center}
\includegraphics[height=8.5cm]{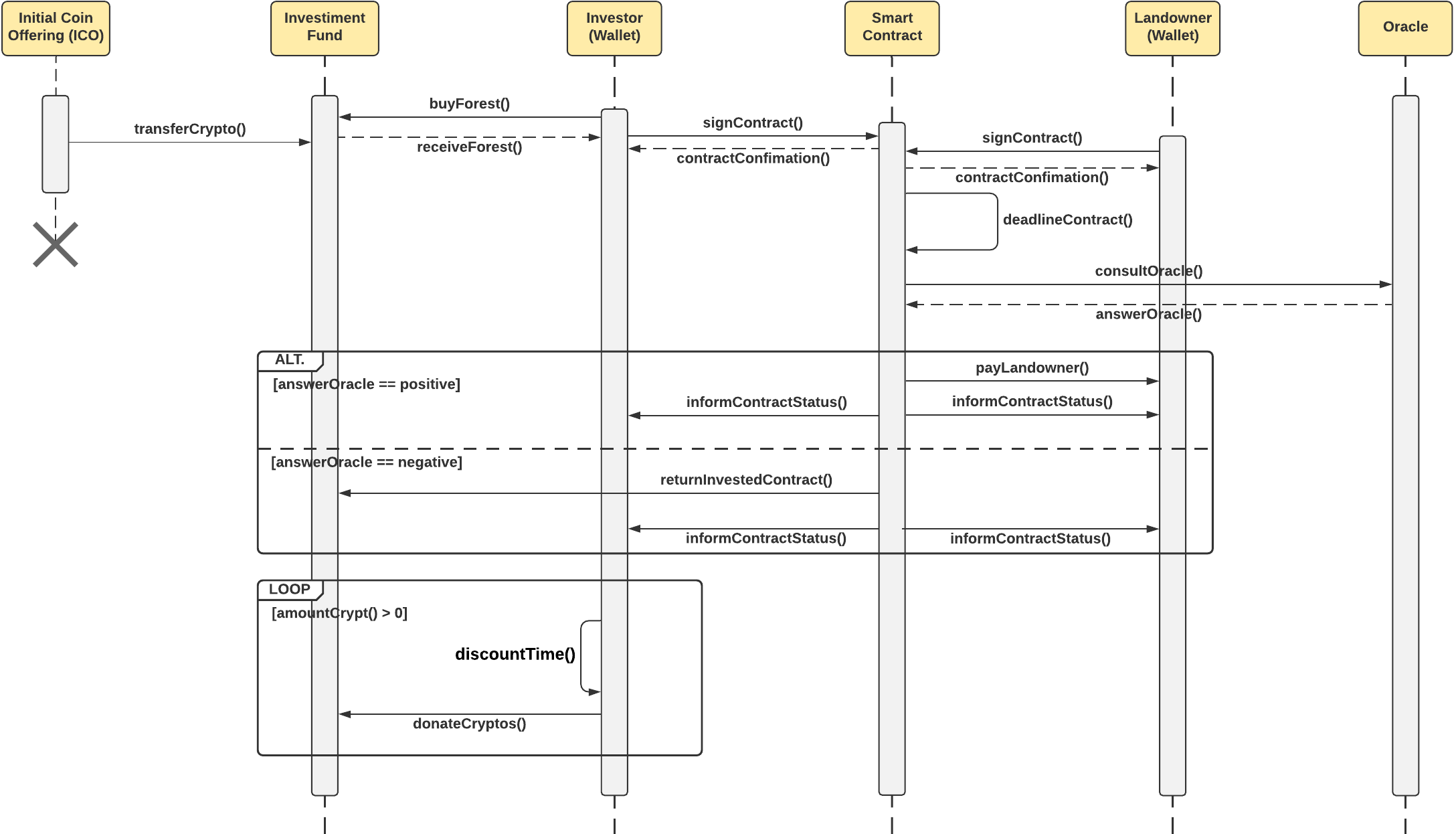}
\end{center}
\end{figure*}

Nesta seção será apresentado o estado atual da implementação da Standing Forest Coin, que envolve desde a concepção dos diagramas de planejamento da moeda ao desenvolvimento dos contratos inteligentes em uma rede de testes da Ethereum.

\subsection{Diagrama de Casos de Uso}

\begin{figure}[h]
\centering
\includegraphics[height=9cm]{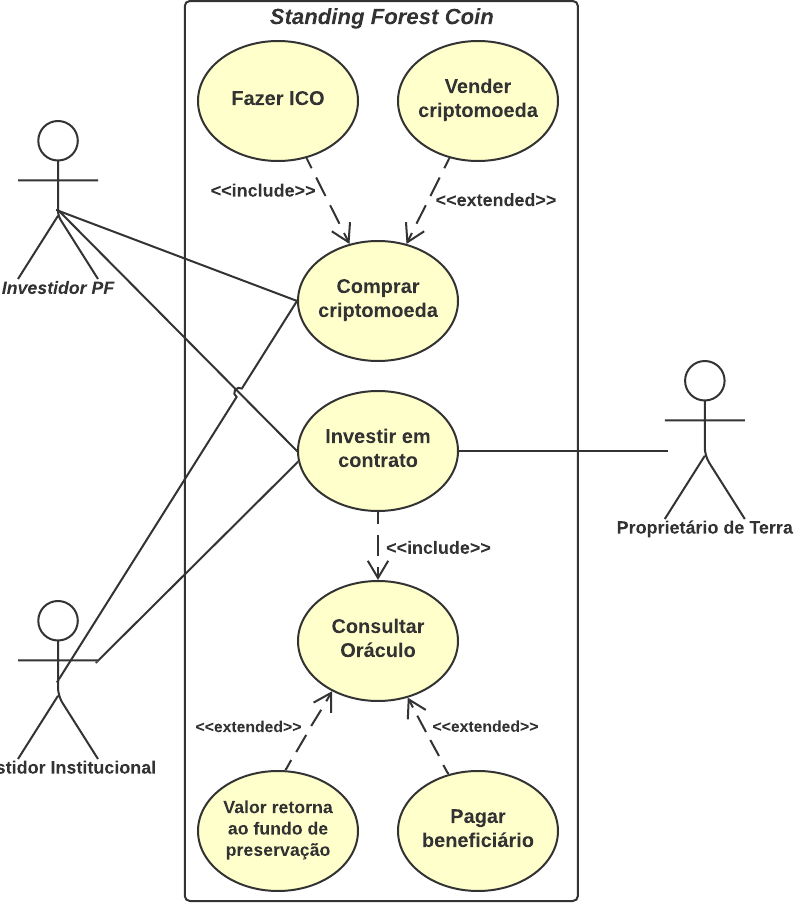}
\caption{{} Caso de Uso de Compra e Investimento em Contrato}
\label{fig:UseCaseDiagram-Case01}
\end{figure}

O diagrama de casos de uso da Figura \ref{fig:UseCaseDiagram-Case01} representa o cenário no qual um investidor, seja ele pessoa física ou um fundo de investimentos, compra SFC e escolhe em qual contrato (proprietário de terra) investirá. Ao término do contrato, o oráculo será consultado pelo contrato para que o beneficiário seja pago ou as criptomoedas sejam transferidas para gestão do fundo de preservação.

\begin{figure}[h]
\centering
\includegraphics[height=4.7cm]{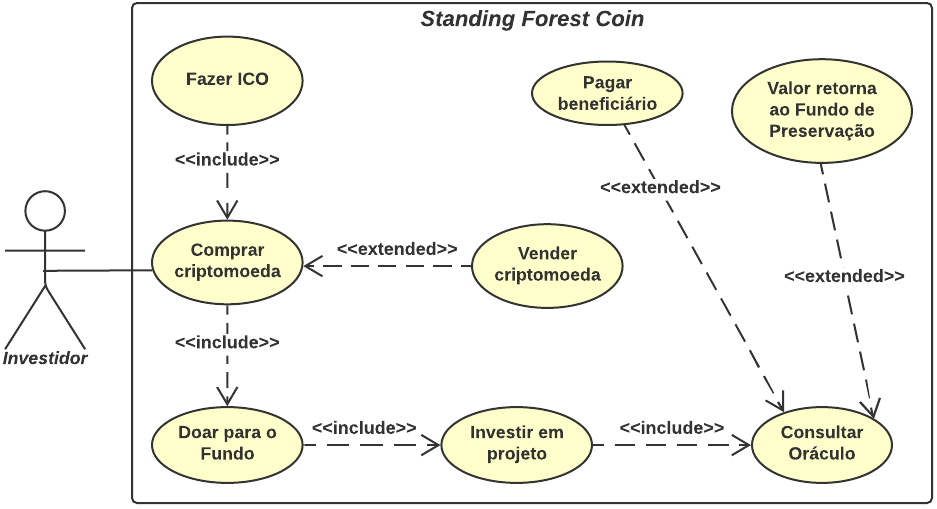}
\caption{{} Caso de Uso de Compra e Doação ao Fundo}
\label{fig:UseCaseDiagram-Case02}
\end{figure}

Na Figura \ref{fig:UseCaseDiagram-Case02}, o diagrama de casos de uso retrata outro caso proposto no qual um investidor compra uma certa quantidade de SFC e decide não investir em nenhum contrato específico. Como a criptomoeda proposta tem o propósito de ser uma ferramenta de doação, a cada ano é descontado 5\% do valor total da carteira do investidor para ser doado para fundo de preservação. 

Dessa forma, o fundo assume a responsabilidade de gestão dos recursos e deve escolher como e em qual contrato inteligente o dinheiro do fundo deve ser investido da mesma forma como acontece no caso de uso da Figura \ref{fig:UseCaseDiagram-Case01}.

\subsection{Diagrama de Sequência}

O diagrama de sequência apresentado na Figura 1 representa os fluxos presentes na implementação da SFC propostos a partir dos casos de usos.

O primeiro fluxo representa o ICO (Initial Coin Offering) da criptomoeda SFC, a partir do ICO o valor será transferido para a carteira do Fundo de Investimento de Preservação que representa a carteira zero para a SFC. A partir da carteira "gênesis", qualquer investidor em qualquer lugar do mundo poderá comprar determinado valor em SFC. Com suas criptomoedas em sua carteira, ele decidirá se irá investir em um contrato de preservação específico ou se seus tokens serão doados anualmente ao Fundo de preservação. 

Quando um contrato de preservação é definido entre as duas partes, no caso o investidor e o guardião da terra, o valor definido pelo investidor é transferido para uma terceira carteira que representa o contrato inteligente. Após o vencimento do contrato, o oráculo será consultado para que se saiba se a preservação da terra foi ou não respeitada. Em caso positivo, o valor do contrato vai para a carteira do guardião da terra e em caso negativo para a carteira inicial do Fundo de Preservação.

\subsection{Implementação}

Como passo inicial para implementação do projeto, foram escritos os contratos inteligentes da moeda SFC utilizando a IDE Remix, que permite desenvolver, compilar e depurar códigos na linguagem de programação Solidity \cite{bouchefra2018}.

A Solidity é uma linguagem de programação de alto nível, influenciada por outras linguagens populares como Python e JavaScript, e que foi desenvolvida com o objetivo de permitir a implementação de \textit{smart contracts} e que tem como alvo a Máquina Virtual Ethereum (EVM) \cite{solidity}, permitindo assim a construção de contratos para a rede Ethereum. Os contratos da SFC foram desenvolvidos em Solidity, na versão 0.5.17 da linguagem.

Após a escrita dos contratos e a sua compilação usando a IDE, foi realizado o processo de \textit{deployment} destes na rede de testes do Ethereum. Esta etapa foi realizada na própria Remix IDE integrando-a com a carteira MetaMask. 

O MetaMask é uma carteira digital que permite gerenciar várias criptomoedas, por meio de uma extensão nos principais browsers de internet \cite{metamask}. No MetaMask foi criada uma carteira para os testes do projeto utilizando a rede de testes Ropsten e então os contratos foram publicados com uma quantidade de 4,000 moedas para o ICO.

\begin{figure}[h]
\centering
\includegraphics[height=4cm]{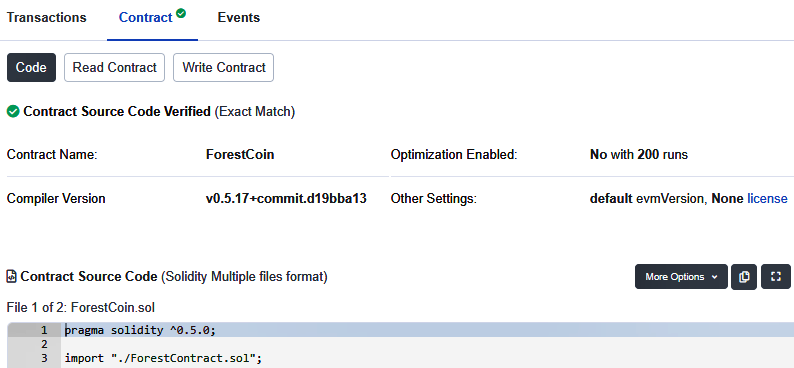}
\caption{{} Verificação dos contratos com o EtherScan}
\label{fig:CheckEtherScan}
\end{figure}

Com o contrato publicado na rede Ethereum, fez-se necessária a verificação dos contratos e dos arquivos fontes do projeto utilizando o serviço EtherScan, como apresentado na Figura \ref{fig:CheckEtherScan}. O EtherScan é um mecanismo de busca que permite pesquisar transações que ocorreram na rede Ethereum \cite{choi2018} e que possibilita a interação direta com os contratos \cite{neuva2020}.

A verificação de um \textit{smart contract}, bem como dos arquivos fonte, é uma etapa importante para garantir que o código dos contratos que foram publicados são exatamente os mesmos que foram desenvolvidos, e permitindo que o contrato seja lido e auditado publicamente por qualquer indivíduo \cite{choi2018}.

\begin{figure}[h]
\centering
\includegraphics[width=8cm]{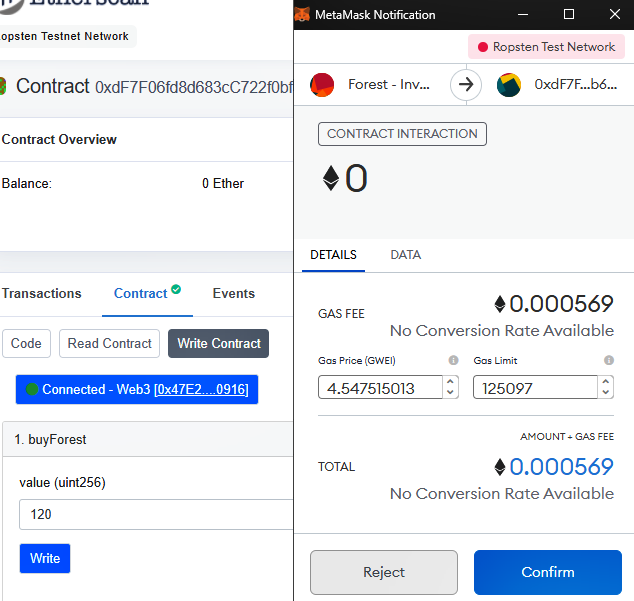}
\caption{{} Aquisição de SFC utilizando o MetaMask}
\label{fig:FirstInteractionContract}
\end{figure}

Após os contratos e o código fonte verificados, torna-se possível a interação com as funções de leitura/escrita do contrato utilizando o EtherScan. Para isso é necessário, além de estar com o contrato aberto, também fazer a conexão ao EtherScan com um gerenciador de carteira digital, no nosso caso, o MetaMask. Na Figura \ref{fig:FirstInteractionContract}, é realizada a primeira interação com o contrato publicado, a compra de 120 tokens da carteira do Fundo de Investimentos diretamente para a carteira de um investidor.

Com a compra dos tokens, foi realizada a simulação da criação de um contrato de dono da terra (contrato esse que receberá os investimentos), tomando como entrada o endereço da carteira do dono da terra. Depois do contrato criado, o EtherScan exibe o endereço do contrato criado, que pode ser usado por exemplo pelo dono da terra para requerer investimentos. Uma vez gerado o contrato, ele permanece público na rede Ethereum.

Após a geração do contrato, foi realizada a transferência de SFCs entre a carteira de um investidor e o contrato do dono da terra, tendo sido transferidos 50 SFC. Por meio do EtherScan, o investidor recebe uma confirmação da doação e pode acompanhar o estado do contrato usando o endereço do contrato do dono da terra.

\begin{figure}[h]
\centering
\includegraphics[width=8.5cm]{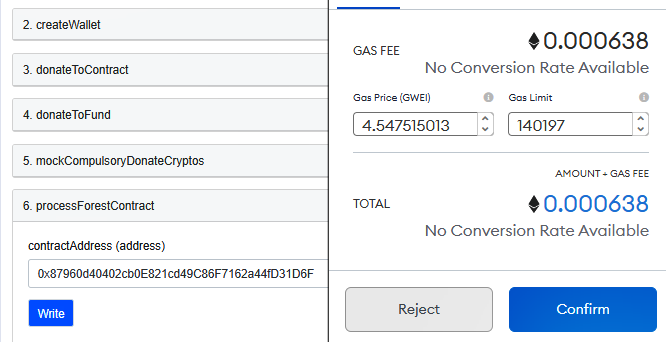}
\caption{{} Processamento do Contrato do Dono da Terra}
\label{fig:ContractProcessing}
\end{figure}

A Figura \ref{fig:ContractProcessing} apresenta o processamento do contrato do dono da terra. O contrato, através de consulta ao oráculo, verifica se a área foi de fato preservada para que seja liberado ou não o valor investido no contrato. Oráculos, em blockchain, são serviços que fornecem informações externas aos \textit{smart contracts}, atuando como uma ``ponte'' entre a blockchain e o mundo exterior \cite{binance2021}. Neste, o oráculo foi implementado como uma simulação, apenas com o objetivo de demonstrar o funcionamento real do contrato.

\begin{figure}[h]
\centering
\includegraphics[width=7cm]{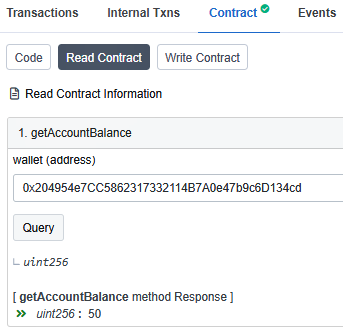}
\caption{{} Resultado do Processamento do Contrato}
\label{fig:ProcessingResult}
\end{figure}

Na Figura \ref{fig:ProcessingResult}, é exibido o resultado do processamento do contrato, onde o dono da terra conseguiu cumprir com o acordo de preservação segundo o oráculo e recebeu o valor investido no contrato, no caso 50 SFCs. 

O contrato de dono da terra, após a validação e movimentação dos tokens, se torna inativo. As transações realizadas com SFC (verificações, processamentos, consultas ao oráculo etc.) são mantidas todas na rede Ethereum, podendo ser visualizadas por qualquer mecanismo de busca, como o EtherScan, mantendo assim a transparência nas operações realizadas.

\section{Conclusão e Próximos Passos}

A blockchain tem se mostrado uma tecnologia muito promissora na questão da transparência dos recursos movimentados durante os processos de criação, validação e transferências entre os contratos inteligentes, servindo de insumo para pesquisas acerca da efetividade na contribuição para o desenvolvimento do país. O aprofundamento no entendimento da tecnologia e a adoção crescente da tecnologia blockchain permitirá uma aceitação maior por partes dos usuários promovendo cada vez mais compartilhamento de código, diminuindo desta forma os custos de transações de troca de valores.

O desenvolvimento até o momento contempla as versões iniciais dos contratos inteligentes da SFC. O foco inicial do desenvolvimento foram as funcionalidades de transferência dos tokens de acordo com os cenários definidos pelos diagramas de caso de uso criados que têm como finalidade promover o financiamento e execução de projetos voltados à proteção do ecossistema amazônico, com mecanismos para assegurar o cumprimento desses projetos através do conceito de Floresta 4.0.

Existem vários passos futuros para o projeto, como por exemplo tornar a experiência do usuário simples. Os conceitos e ferramentas de uso não estão massificados na sociedade. No caso estudado, por exemplo, os usuários que enviam transações para a rede Ethereum precisam instalar o MetaMask e ter Ether para pagar a taxa de encargo da blockchain.

Outro ponto crucial é conseguir financiamento necessário para uma carga inicial que possa dar lastreamento à criptomoeda criada, além de fazer os interessados comprarem a ideia da SFC. Também é preciso reavaliar constantemente se o Ethereum é realmente a plataforma de blockchain mais adequada para a solução. É necessário acompanhar o dinâmico desenvolvimento e amadurecimento do mercado de blockchain considerando os requisitos da solução, especialmente privacidade dos dados.

\section*{Agradecimentos}

Os autores agradecem à Coordenação de Aperfeiçoamento de Pessoal de Nível Superior (CAPES) e à Fundação de Amparo à Pesquisa do Estado de São Paulo (FAPESP) pelo apoio ao desenvolvimento parcial deste projeto.

\bibliographystyle{ieeetr}  

\bibliography{References}

\ifCLASSOPTIONcaptionsoff
  \newpage
\fi


%

\begin{IEEEbiography}[{\includegraphics[width=1in,height=1.25in,clip]{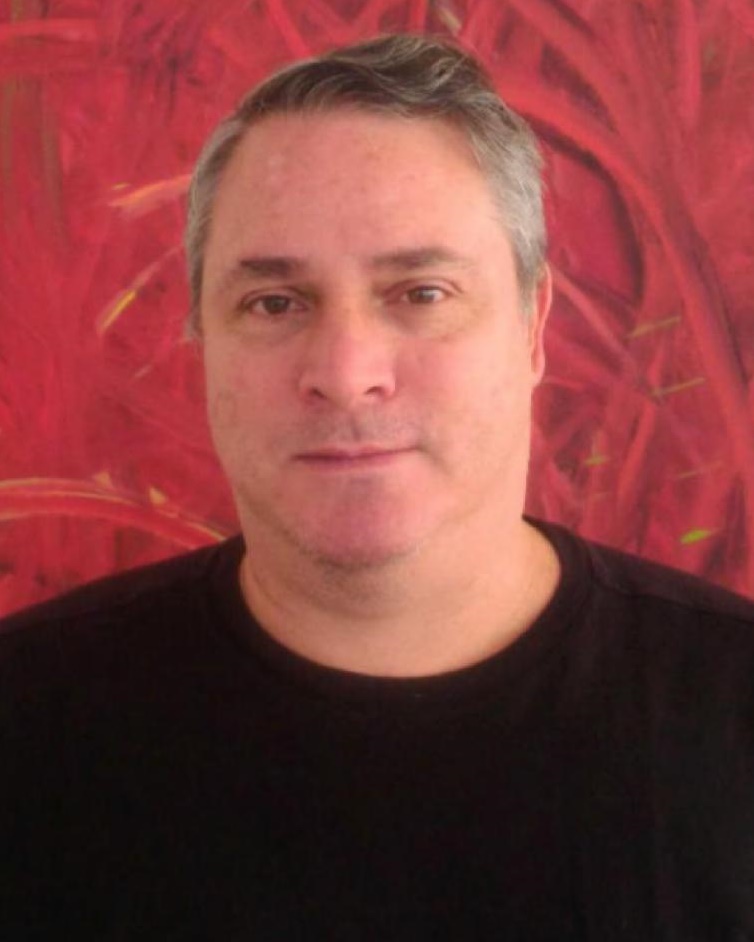}}]{Marcelo de Abreu Borges}
foi sócio fundador da Tudor Asset Management, umas das primeiras gestoras de recursos independentes, na década de 1990. No setor educacional foi um dos sócios fundadores da ABECE S/A, grupo que em 2009 incorpora a rede de escolas e faculdade Damásio de Jesus, formando a Damásio Educacional S/A, vendida em 2015 para a Devry Brasil, subsidiária do grupo educacional americano Devry. Mestre em Análise de Políticas Internacionais pela PUC Rio, é CEO da Wisecare Tecnologia e pesquisador interdisciplinar.
\end{IEEEbiography}

\begin{IEEEbiography}[{\includegraphics[width=1in,height=1.25in,clip,keepaspectratio]{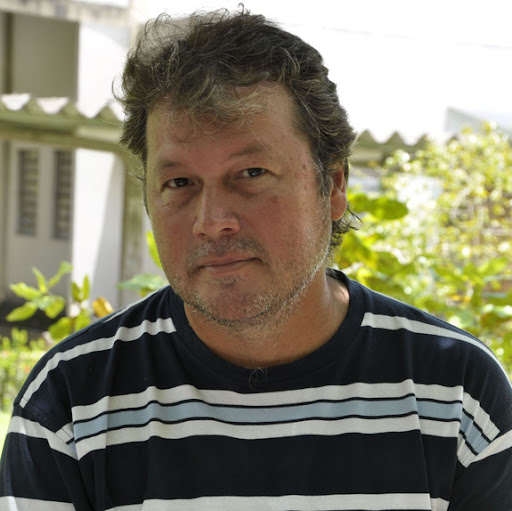}}]{Guido Lemos de Souza Filho}
é professor Titular do Centro de Informática na Universidade Federal da Paraíba (UFPB) e pesquisador do Laboratório de Aplicações de Vídeo Digital (LAVID). Foi um dos desenvolvedores do middleware Ginga, adotado como padrão no Sistema Brasileiro de Televisão Digital e de vários outros países da América Latina e Africa. Desenvolve redes de distribuição de vídeo, acessibilidade e aplicações de blockchain, servidores de vídeo para transmissão ao vivo e sob demanda (DLive e DVod) e softwares para acessibilidade, como o VLibras (usado nos sites brasil.gov.br, senado.leg.br e câmara.leg.br). Atua também como membro do Conselho Deliberativo do Fórum do Sistema Brasileiro de Televisão Digital. Recebeu vários prêmios nacionais e internacionais na área de tecnologia e sociedade, dentre eles o Prêmio Trip Transformadores 2010.
\end{IEEEbiography}

\begin{IEEEbiography}[{\includegraphics[width=1in,height=1.25in,clip]{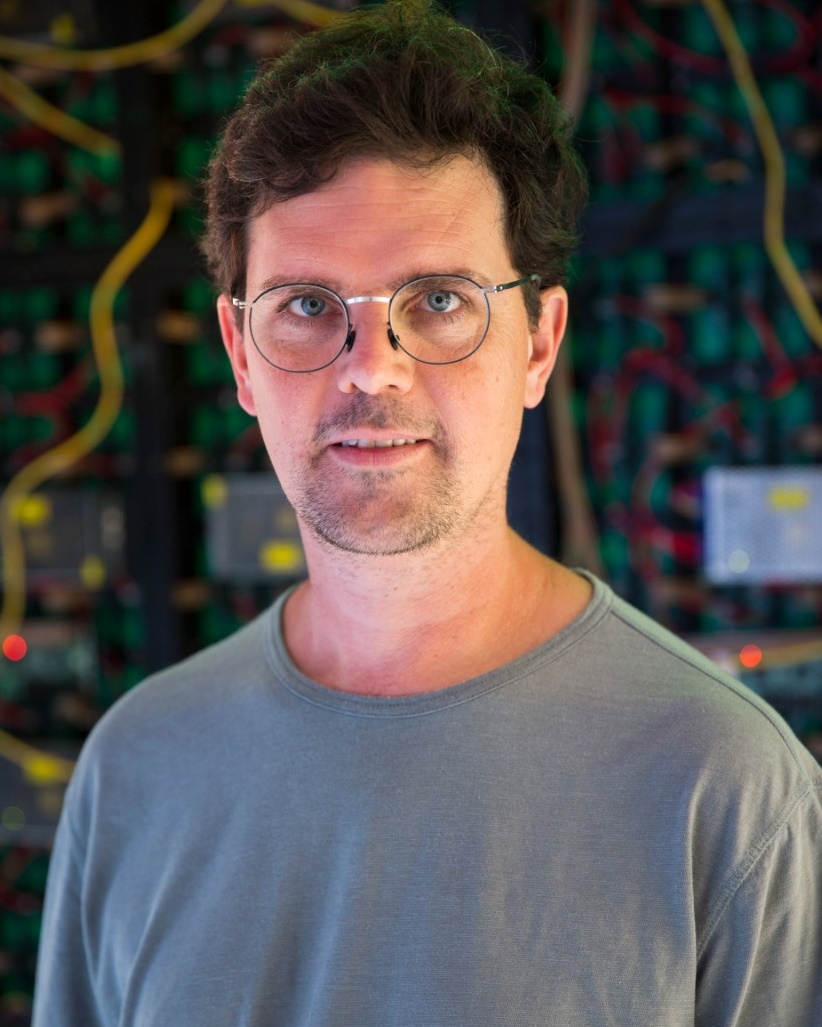}}]{Cicero Inacio da Silva}  atualmente é professor associado da Unifesp na área de mídias computacionais aplicadas à ciência, saúde e cultura. É pesquisador associado ao Cultural Analytics Lab na City University of New York (CUNY)/CALIT2 e membro do Comitê de Avaliação Interdisciplinar da Canada Foundation for Innovation (CFI). Foi professor visitante na Brown University, na University of California e recebeu menção honrosa do Prix Ars Electronica (Áustria) na área de Digital Communities em 2010. 
É um dos organizadores e membro fundador da associação californiana Cinegrid (www.cinegridbr.org).
\end{IEEEbiography}

\begin{IEEEbiography}[{\includegraphics[width=1in,height=1.25in,clip]{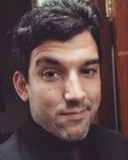}}]{Anderson Marinho Pontes de Barros} tem bacharelado em Ciência da Computação pela Universidade Federal da Paraíba (2016) e atualmente é mestrando em Informática pela Universidade Federal da Paraíba. http://lattes.cnpq.br/6509604951226199
\end{IEEEbiography}

\begin{IEEEbiography}[{\includegraphics[width=1in,height=1.25in,clip,keepaspectratio]{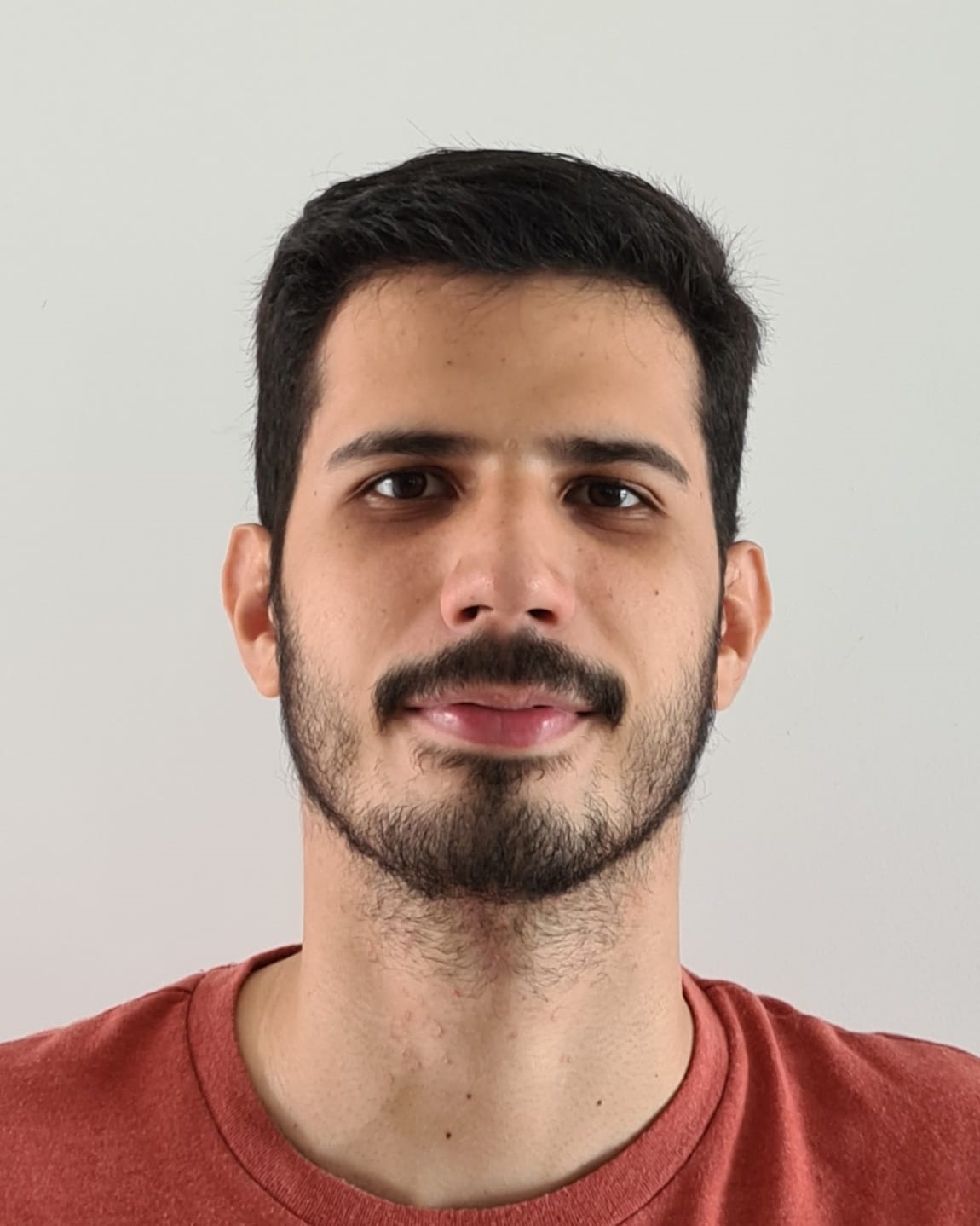}}]{Raul Victor Barreto Jácome Britto} tem bacharelado em Engenharia de Computação pela Universidade Federal da Paraíba (2018) e atualmente é Mestrando em Informática pela Universidade Federal da Paraíba. Possui experiência nas áreas de Big Data e Machine Learning.
\end{IEEEbiography}

\begin{IEEEbiography}[{\includegraphics[width=1in,height=1.25in,clip,keepaspectratio]{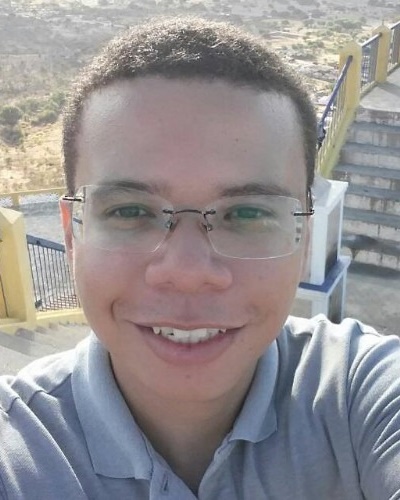}}]{Nivaldo Mariano de Carvalho Junior} é graduado em Gestão da Tecnologia da Informação pelo Centro Universitário de João Pessoa (2017), e atualmente é mestrando em Informática pela Universidade Federal da Paraíba. http://lattes.cnpq.br/0866664264664231
\end{IEEEbiography}

\begin{IEEEbiography}[{\includegraphics[width=1in,height=1.25in,clip]{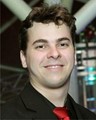}}]{Daniel Faustino Lacerda de Souza} tem bacharelado em Ciência da Computação pela Universidade Federal da Paraíba (2008), mestrado em Informática pela Universidade Federal da Paraíba (2010) e doutorado em Engenharia Elétrica e da Computação pela Universidade Federal do Rio Grande do Norte (2017). http://lattes.cnpq.br/7175882793842898
\end{IEEEbiography}


\end{document}